\documentclass[final,3p,times,12pt,authoryear]{elsarticle}
\usepackage[utf8]{inputenc}
\usepackage{graphicx}
\usepackage{amsmath}
\usepackage{wrapfig}
\usepackage{multicol}
\usepackage{multirow}
\usepackage{bigdelim}
\usepackage[version=4]{mhchem}
\usepackage{siunitx}
\usepackage{longtable,tabularx}
\usepackage{tikz}

\usepackage{amssymb}
\usepackage{amsmath}

\begin{document}

\begin{frontmatter}



\title{A data-driven approach for modeling large-amplitude flow-induced oscillations of elastically mounted pitching wings} 


\author[a,b]{Yuanhang Zhu\corref{ca}}
\cortext[ca]{Corresponding author}
\ead{yuanhang\_zhu@alumni.brown.edu}
\author[a]{Kenneth Breuer} 

\affiliation[a]{organization={Center for Fluid Mechanics, School of Engineering, Brown University},
            city={Providence},
            state={RI},
            postcode={02912}, 
            country={USA}}

\affiliation[b]{organization={Department of Mechanical Engineering, University of California},
            city={Riverside},
            state={CA},
            postcode={92521}, 
            country={USA}}

\begin{abstract}
We propose and validate a data-driven approach for modeling large-amplitude flow-induced oscillations of elastically mounted pitching wings. We first train a neural networks regression model for the nonlinear aerodynamic moment using data obtained from experimental measurements during prescribed pitching oscillations and at fixed angles of attack. We then embed this model into an ordinary differential equation solver to solve the governing equation of the passive aeroelastic system with desired structural parameters. The system dynamics predicted by the proposed data-driven approach are characterized and compared with those obtained from physical experiments. The predicted and experimental pitching amplitude, frequency and aerodynamic moment responses are found to be in excellent agreement. Both the inertia-dominated mode and the hydrodynamic-dominated mode are successfully predicted. The transient growth and saturation of the pitching oscillation amplitude and the aerodynamic moment are also faithfully captured by the proposed approach. Additional test cases demonstrate the broad applicability and good scalability potential of this approach.
\end{abstract}




\end{frontmatter}




\section{Introduction}

The interactions between free-stream flows and elastically supported structures can lead to large-amplitude self-sustained oscillations, such as vortex-induced vibrations of cylinders \citep{williamson2004vortex,gabbai2005overview,williamson2008brief} and aeroelastic instabilities of airfoils \citep{dowell1989modern,dowell2001modeling,dowell2003nonlinear,dimitriadis2009bifurcation,dimitriadis2017introduction}. These flow-induced oscillations can pose severe risks, potentially leading to structural failure in extreme conditions \citep{larsen1998advances}, but they also present opportunities for hydrokinetic energy harvesting from river or tidal flows \citep{xiao2014review,young2014review,onoue2015large,su2019resonant,wang2020state}, providing alternative solutions for renewable energy generation.

\begin{figure}
\centering
\includegraphics[width=.75\textwidth]{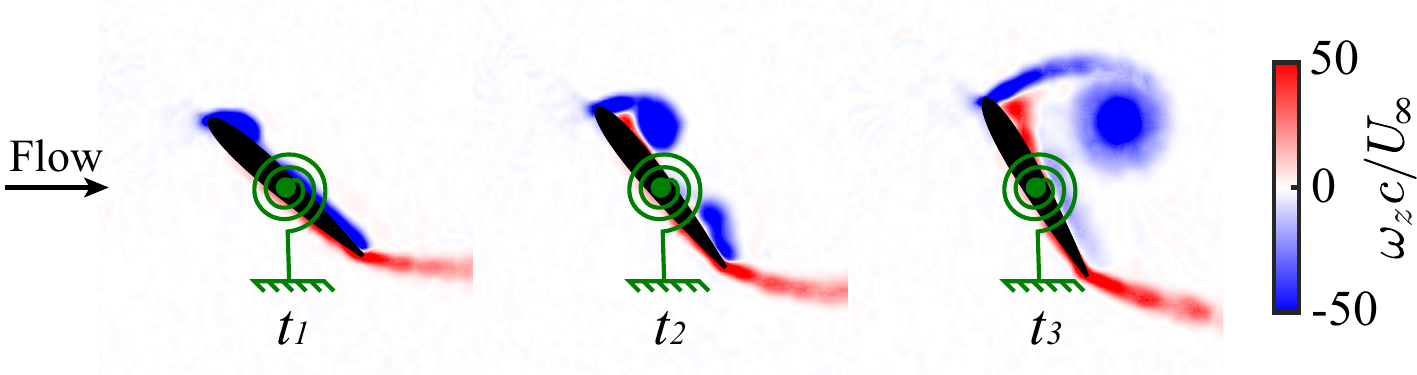}
\caption{The formation, growth, and shedding of a strong leading-edge vortex (LEV) on an elastically mounted pitching wing undergoing large-amplitude flow-induced oscillations. Snapshots are reproduced from particle image velocimetry (PIV) experiments of \citet{zhu2020nonlinear}. Blue and red regions represent negative (clockwise) and positive (counterclockwise) vorticities, respectively. $t_{1-3}$ refers to time instances of the LEV evolution, $\omega_z$ represents spanwise vorticity, $c$ denotes chord length, and $U_\infty$ is the flow speed. Note that these vorticity snapshots are presented solely to illustrate the LEV evolution. No PIV data are otherwise used in the present study.}
\label{fig.LEV}
\end{figure}

To comprehensively understand and effectively control these complex flow-induced oscillations, numerous efforts have been made to study the corresponding fluid-structure interactions (FSI). One common way of simplifying an FSI system with a large number of degrees of freedom (DOFs) is to model it as a spring-mass-damper system subjected to nonlinear fluid forcing \citep{khalak1996dynamics,dugundji2008some}. In the present study, we consider an elastically mounted aeroelastic system with only one degree of freedom in pitching:
\begin{equation}
    I \ddot{\theta} + b \dot{\theta} + k \theta = M,
\label{eqn.govern}
\end{equation}
where $\theta$, $\dot{\theta}$, $\ddot{\theta}$ are pitching angle, rate, acceleration, $I$, $b$, $k$ are wing inertia, damping, stiffness, and $M$ is the pitching aerodynamic moment (for abbreviation, the ``pitching aerodynamic moment'' is hereafter referred to as the ``aerodynamic moment''). In this equation, the left hand side is a simple spring-mass-damper system. For the large-amplitude flow-induced oscillations considered in this study, the unsteady fluid forcing term on the right hand side of Eq. \ref{eqn.govern}, $M$, mainly results from the formation, growth, and shedding of vortices around the wing (particularly the LEV, see Fig. \ref{fig.LEV}). This unsteady pitching aerodynamic moment is often found to be a highly nonlinear function of the pitching angle, rate, and acceleration, which is difficult to model and predict. Several other parameters, including the Reynolds number, the wing geometry, and the surface roughness, may also be important. Therefore, to model and predict the dynamics of the aeroelastic system described in Eq. \ref{eqn.govern}, it is critical to develop a cost effective yet accurate model for the unsteady aerodynamic moment term.

Classic models from \citet{theodorsen1935general} and \citet{garrick1936propulsion} provide an analytical solution for estimating the unsteady aerodynamic loads on two-dimensional airfoils undergoing harmonic oscillations. However, these linear models are based on potential flow theory assuming a thin airfoil, and are only applicable to attached flows and small-amplitude oscillations. Although these inviscid models account for unsteady vortex sheddings, they use a vorticity sheet that flows downstream along a line to represent the wake. To what extent do these models apply to highly separated, vortex-dominated flows and large-amplitude oscillations remains an active research direction \citep{mccroskey1982unsteady,baik2012unsteady,liu2015unsteady,otomo2021unsteady}.

Discrete vortex-based models have been proposed thereafter to incorporate the nonlinear and unsteady effects of discrete shed vortices \citep{clements1975representation,katz1981discrete,xia2013lift,wang2013low,ramesh2013unsteady,ford2013lift,yan2014geometrically,riso2016nonlinear,onoue2016vortex}. See also the review by \citet{eldredge2019leading}. These models provide valuable tools for predicting aerodynamic forces in highly separated flows, capturing the essential physics with significantly lower computational cost than full numerical simulations. Moreover, they have good physical interpretability, in that the unsteady aerodynamic load can be associated with the corresponding flow physics, such as variations of leading-edge suction and vortex shedding. However, these two-dimensional, potential flow-based models neglect nonlinearities associated with three-dimensional vortex shedding and viscous effects. As a result, they cannot accurately capture the flow physics whenever these nonlinearities are significant. Some models also require additional inputs, such as the separation location on the airfoil \citep{ramesh2013unsteady}, and vortex position and strength \citep{ford2013lift,onoue2016vortex}. Therefore, careful validation against experimental or high-fidelity numerical data is essential.

Modeling the two-way coupling between the fluid and the structure using high-fidelity numerical simulations (e.g. \citet{goza2017strongly,menon2019flow}) is nontrivial and computationally expensive \citep{henshaw2007non}, and not practical for instantaneous aeroelastic predictions. Similarly, conducting two-way coupling FSI experiments can also be challenging. For example, to study the effect of inertia and stiffness on the flow-induced oscillations of a pitching airfoil at a fixed Reynolds number, one would need to physically swap airfoils with different masses and change torsional springs with different stiffness coefficients. An alternative approach is to use a cyber-physical system, where the structural parameters can be simulated by a force-feedback control loop \citep{mackowski2011developing,onoue2016vortex,zhu2020nonlinear}. However, the implementation of cyber-physical systems usually requires careful design of the feedback control loop, special attention to hardware connections for reducing noise, and adequate computational power for instantaneous sampling and processing, making the experiments expensive and less accessible.

\begin{wrapfigure}{l}{0.45\textwidth}
\centering
\includegraphics[width=0.45\textwidth]{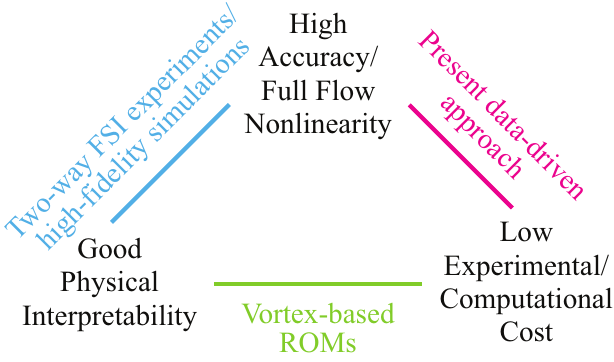}
\caption{Trade-offs between high accuracy, good interpretability, and low experimental/computational cost in modeling large-amplitude aeroelastic oscillations.}
\label{fig.tradeoff}
\end{wrapfigure}

The trade-offs between high accuracy, good interpretability, and low experimental and computational cost in various approaches to modeling large-amplitude aeroelastic oscillations are summarized in Fig. \ref{fig.tradeoff}. As we discussed earlier, vortex-based reduced-order models (ROMs) feature good physical interpretability and low computational cost. High-fidelity two-way coupling simulations and experiments also have the advantage of good interpretability, as the simulated or measured flow physics results from physics-based numerical models or physical experiments. As these approaches can usually capture the full flow dynamics and nonlinearities, they also achieve high accuracy. The goal of the present study is to develop a new approach that can accurately model large-amplitude flow-induced oscillations of elastically mounted pitching wings with full nonlinearity and at a low experimental/computational cost. To achieve this goal, we propose a data-driven approach that uses data obtained from prescribed (one-way coupling) experiments to train a machine-learning regression model of the unsteady aerodynamic pitching moment as a function of the pitching angle, rate, and acceleration. We then combine the trained model with the aeroelastic governing equations and simulate the instantaneous aeroelastic dynamics (two-way coupling) by solving the governing equations based on the desired structural parameters. We use one-way coupling FSI to train the machine-learning model because it is substantially easier and more cost effective to implement in experiments and simulations compared to two-way coupling FSI, as it only requires prescribing structural motions and measuring or simulating the fluid response.

Previous studies have shown that large-amplitude flow-induced oscillations are largely sinusoidal, and based on this observation, an energy method has been used to accurately characterize two-way aeroelastic oscillations using one-way prescribed oscillations \citep{morse2009prediction,bhat2013stall,menon2019flow,zhu2020nonlinear}. The energy method uses prescribed sinusoidal motions (one-way coupling) to map out the cycle-averaged energy transfer between the ambient fluid and the elastic structure over a large range of oscillating frequencies and amplitudes. The stability of the passive aeroelastic system (two-way coupling) can thus be inferred from the energy transfer, as the energy injected by the ambient fluid should always balance the energy dissipated by the structural damping for self-sustained oscillations. However, this approach is only capable of predicting the \emph{overall} stability of passive systems and the \emph{cycle-averaged} amplitude and frequency of self-sustained flow-induced oscillations. It is not able to make predictions for the \emph{instantaneous} dynamics of the system, such as the bifurcation points, the transitional dynamics, and the mode of oscillations, because the energy transfer is evaluated in a cycle-averaged manner.

There have also been CFD-based studies that focus on using one-way coupling prescribed motions in only a small number of dominant modes to reconstruct ROMs for analyzing aeroelastic instabilities \citep{raveh2001efficient,lucia2004reduced,silva2004development,henshaw2007non,afonso2017review,torregrosa2021application,simiriotis2023numerical}. However, these ROMs typically assume small-amplitude motions near an equilibrium solution and, therefore, do not capture full nonlinearities in the flow, and might not be applicable for modeling large-amplitude flow-induced oscillations, such as the one depicted in Fig. \ref{fig.LEV}. Hence, using one-way coupling \emph{experimental data} to construct fully nonlinear aeroelastic models is of significant value. Examples of these models include studies of sloshing-induced aeroelastic instabilities \citep{saltari2022sloshing,pizzoli2023neural}, where experiment-based models have demonstrated the ability to capture nonlinear phenomena such as the onset of limit-cycle oscillations. However, to the best of our knowledge, this approach has not yet been applied to modeling large-amplitude flow-induced oscillations in vortex-dominated flows.

With the rapid development of machine learning tools in fluid mechanics \citep{brunton2016discovering,brenner2019perspective,raissi2019physics,brunton2020machine,kou2021data,menon2021quantitative,lee2021reduced,lee2022leading,siddiqui2022data,mishra2023predicting,ribeiro2023machine,hickner2023data,fukami2023grasping,graff2023information,chen2023sparse,carter2023low,le2023improving,beregi2023using}, we see a possibility of better bridging one-way and two-way coupling FSI problems and substantially reducing the experimental and computational costs of modeling nonlinear aeroelastic systems while capturing full nonlinearities. As summarized in Fig. \ref{fig.tradeoff}, our goal is to develop a less experimentally and computationally expensive approach to model the two-way coupling FSI of large-amplitude flow-induced oscillations. Despite its low interpretability, our approach is less expensive than full FSI experiments and simulations, while being more accurate and incorporating higher nonlinearities compared to physics-based ROMs.

In the following sections, we will first introduce our methodology (\S \ref{sec.method}), including the experimental and machine learning setup (\S \ref{sec.expts}), as well as the data assembling (\S \ref{sec.data_prepare}), training (\S \ref{sec.training}), and validation (\S \ref{sec.prediction}) steps. This is followed by comparing the results obtained from the data-driven approach with fully two-way aeroelastic experimental measurements (\S \ref{sec.results}). Finally, we discuss the results and summarize all the key findings in \S \ref{sec.conclusion}.

\section{Methods}{\label{sec.method}}

\subsection{Experimental setup and machine learning workflow}{\label{sec.expts}}

\begin{figure}
\centering
\includegraphics[width=.85\textwidth]{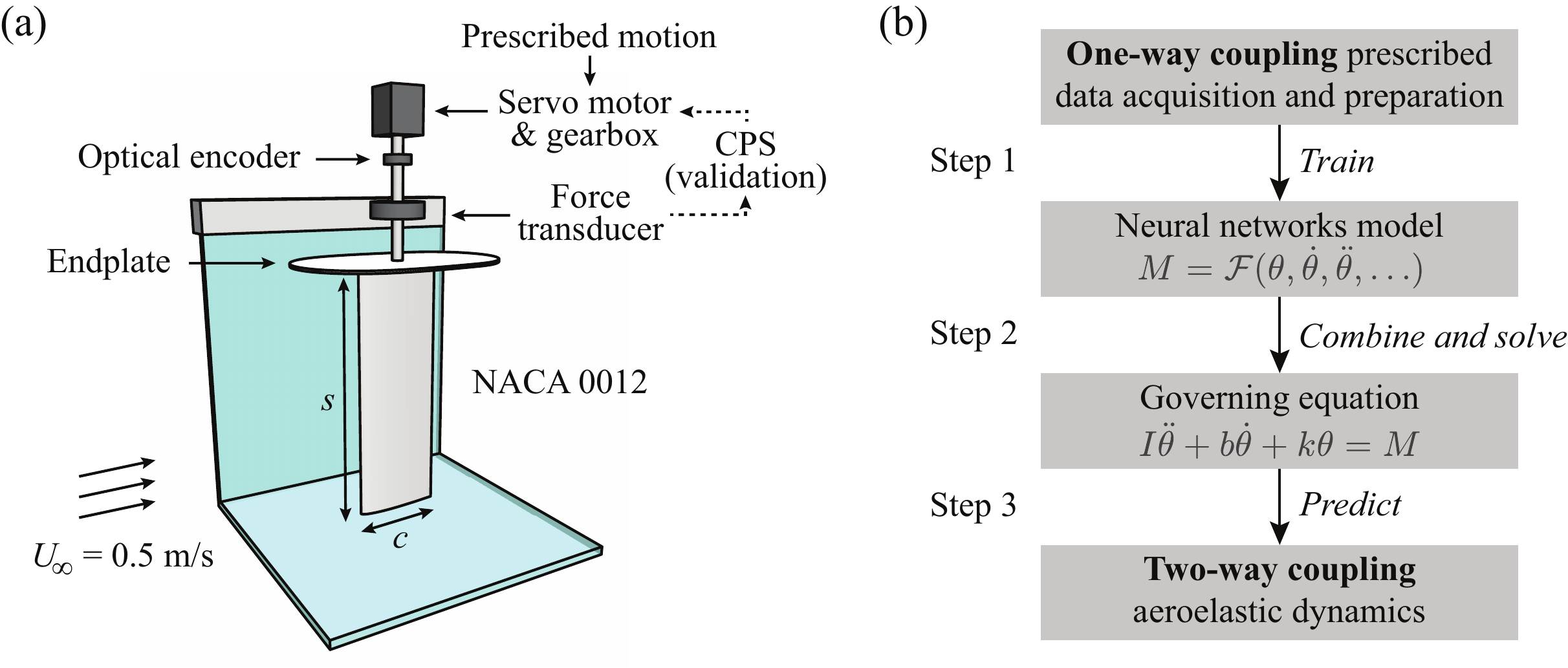}
\caption{(a) A schematic of the experimental setup. (b) A flow chart of the proposed data-driven approach.}
\label{fig.setup}
\end{figure}

As we discussed in the previous section, the goal of the present work is to propose a data-driven approach to bridge one-way and two-way coupling fluid-structure interaction problems and use the approach to model large-amplitude aeroelastic oscillations. To achieve this goal, we first conduct prescribed-motion experiments to obtain the training data. A schematic of the experimental setup is shown in Fig. \ref{fig.setup}(a). We mount a NACA 0012 wing vertically in a circulating free-surface water tunnel (test section $W \times D \times L = 0.8~\mathrm{m} \times 0.6~\mathrm{m} \times 4.0~\mathrm{m}$). The wing has a chord length of $c=0.1$ m and a span of $s=0.3$ m. An endplate is attached to the wing root to skim surface waves and eliminate tip vortices. The endplate also acts as a symmetry plane, resulting in an effective aspect ratio of $AR=6$. As we will show later, the specific geometry of the wing (e.g. pitching axis, cross-section shape, aspect ratio, sweep angle, endplates, etc) does not affect the effectiveness of the proposed data-driven approach. The free-stream flow velocity is fixed at $U_{\infty} = 0.5$ m/s, which results in a chord-based Reynolds number of $Re \equiv \rho U_{\infty} c / \mu = 50,000$, where $\rho$ and $\mu$ are water density and dynamic viscosity, respectively.

The wing pitches around a vertical shaft at the mid chord, which is further connected to a six-axis force/moment transducer (ATI 9105-TIF-Delta-IP65), an optical encoder (US Digital E3-2500), and a servo motor (Parker SM233AE) coupled with a 5:1 gearbox (Automation Direct PGCN23-0525). The force/moment transducer measures the aerodynamic moment, $M$, experienced by the wing and the encoder measures the pitching angle, $\theta$. Both signals are recorded on a desktop computer via a data acquisition (DAQ) board (National Instruments PCIe-6353) at a rate of 1,000 Hz. Note that all the training data acquired for the proposed data-driven approach are from \emph{prescribed} pitching motions or \emph{static} moment measurements (one-way coupling).

Fig. \ref{fig.setup}(b) shows a flow chart of the proposed data-driven approach. The first step is to acquire and prepare the one-way coupling training data, which will be discussed in detail in the following subsection (\S \ref{sec.data_prepare}). We then use the prepared data to train a neural networks (NN) model (\S \ref{sec.training}) for the unsteady aerodynamic moment as a function of the pitching angle, rate, and acceleration. The trained model is later combined with the governing equation to make predictions of the dynamics of the two-way coupling aeroelastic system (\S \ref{sec.prediction}).

\subsection{Training data acquisition and preparation}{\label{sec.data_prepare}}

\begin{figure}
\centering
\includegraphics[width=.8\textwidth]{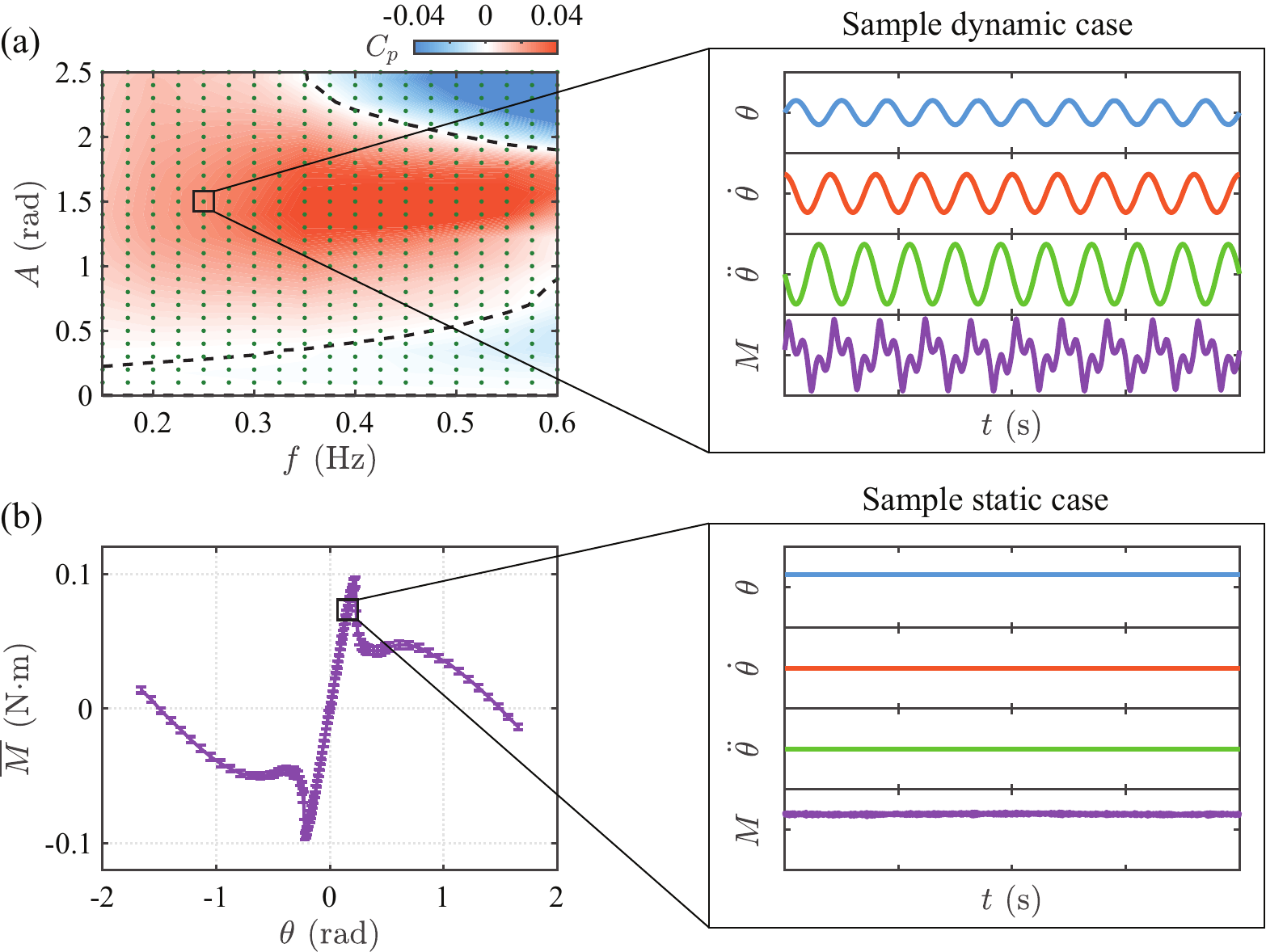}
\caption{(a) Power transfer map of the system, generated by one-way coupling prescribed sinusoidal motions. Green dots represent the locations of the $i=475$ data points. Zoom-in window: time series data of the three predictors ($\theta,\dot{\theta}$ and $\ddot{\theta}$) and the response ($M$) for case $A=1.5$ and $f=0.25$ Hz. (b) Static aerodynamic moment measurement as a function of the pitching angle ($j=105$ cases). Zoom-in window: time series of the training data $\theta,\dot{\theta},\ddot{\theta}$ and $M$ for case $\theta=0.157$.}
\label{fig.data}
\end{figure}

The one-way coupling training data contains two parts. The first part is the \emph{dynamic} data acquired from prescribed sinusoidal pitching oscillations and the second part is the \emph{static} data obtained from aerodynamic moment measurements at fixed angles of attack. Fig. \ref{fig.data}(a) and (b) illustrate how the training data is extracted from the dynamic and static measurements, respectively.

To obtain the dynamic training data, we prescribe sinusoidal pitching motions of the wing using the following equation \citep{zhu2020nonlinear}
\begin{equation}
    \theta = A \sin (2\pi f_p t),
    \label{eqn.sine_motion}
\end{equation}
where $A$ is the pitching amplitude, $f_p$ is the pitching frequency and $t$ is the time. $A$ is varied from 0 to 2.5 rad with a step size of 0.1 rad and $f$ is varied from 0.15 Hz to 0.6 Hz with a step size of 0.025 Hz, resulting in $i=475$ total prescribed (dynamic) cases. These cases were previously used to construct the power transfer map in \citet{zhu2020nonlinear}. Here we replot the power transfer map in Fig. \ref{fig.data}(a) without considering structural damping. The power transfer coefficient is calculated as
\begin{equation}\label{eqn.power}
    C_p = \frac{f_p^*}{n} \int_{t_0}^{t_0+nT} C_M \dot{\theta}^*~d t^*,
\end{equation}
where $f_p^*=f_p c/U_\infty$ is the non-dimensional pitching frequency, $n=10$ is the number of pitching cycles, $t_0$ is the starting time, $T=1/f_p$ is the pitching period, $C_M$ is the moment coefficient, $\dot{\theta}^*=\dot{\theta}c/U_{\infty}$ is the non-dimensional pitching rate, and $t^*=t U_{\infty}/c$ is the non-dimensional time. On this power transfer map, orange regions represent $C_p>0$, where more power is being injected into the system from the ambient flow, whereas blue regions represent $C_p<0$, where more power is being damped into the ambient flow from the system. The black dashed curve represents neutral power transfer, $C_p=0$, where the system is at an equilibrium state. The green dots on the map represent the locations of the $i=475$ dynamic cases. The zoom-in window shows the time series data of the pitching angle, $\theta$, the pitching rate, $\dot{\theta}$, and the pitching acceleration, $\ddot{\theta}$, and the aerodynamic moment, $M$, measured at the sample case $A=1.5$ and $f=0.25$ Hz. The pitching rate and the pitching acceleration are derived from the time-resolved pitching angle. For each prescribed dynamic case, 10 pitching cycles are recorded.

In addition to the dynamic data, the static data is also important for predicting the dynamic of the system, in particular the onset of flow-induced oscillations \citep{zhu2020nonlinear,zhu2023flow}. To obtain the static training data, we measure the static pitching moment when the wing is fixed at different angles of attack for $-1.66~\mathrm{rad} \leq \theta \leq 1.66~\mathrm{rad}$ with an adaptive step size (0.009 rad near $\theta=0$ and 0.09 rad near $\theta=\pm 1.66$ rad). The static moment curve is shown in Fig. \ref{fig.data}(b). The resultant pitching rate and pitching acceleration for these static cases are both zero. The zoom-in window shows the corresponding time series data for the sample static case $\theta=0.157$. In total, $j=105$ static cases are measured and each measurement lasts 20 seconds. For both the dynamic and the static cases, no filter is applied to the aerodynamic moment data. The effect of noise filtering on the performance of the trained machine-learning model is further discussed in \ref{sec.noise}.

\begin{table}[ht!]
\centering
\begin{tabular}{c|c c c|c|c|ll}
\multirow{18}{*}[-1em]{Predictors:} & $\theta_{D11}$ & $\dot{\theta}_{D11}$ & $\ddot{\theta}_{D11}$ & \multirow{18}{*}[-1em]{~~~~~Response:} & $M_{D11}$ & \multirow{4}{*}[1.75em]{\rdelim\}{4}{*}[~Dynamic case 1]} & \\
& $\theta_{D12}$ & $\dot{\theta}_{D12}$ & $\ddot{\theta}_{D12}$ & & $M_{D12}$ & \\
& $\theta_{D13}$ & $\dot{\theta}_{D13}$ & $\ddot{\theta}_{D13}$ & & $M_{D13}$ & \\
& $\vdots$ & $\vdots$ & $\vdots$  & & $\vdots$ & \\
& $\theta_{D21}$ & $\dot{\theta}_{D21}$ & $\ddot{\theta}_{D21}$  & & $M_{D21}$ & \multirow{4}{*}[1.75em]{\rdelim\}{4}{*}[~Dynamic case 2]}\\
& $\theta_{D22}$ & $\dot{\theta}_{D22}$ & $\ddot{\theta}_{D22}$  & & $M_{D22}$ & \\
& $\theta_{D23}$ & $\dot{\theta}_{D23}$ & $\ddot{\theta}_{D23}$  & & $M_{D23}$ & \\
& $\vdots$ & $\vdots$ & $\vdots$  & & $\vdots$ & \\
& \multicolumn{3}{c|}{$\cdots$} & & $\cdots$ & \hspace{0.9em}Dynamic case 3 to $i$ \\
& $\theta_{S11}$ & 0 & 0  & & $M_{S11}$ & \multirow{4}{*}[1.75em]{\rdelim\}{4}{*}[~Static case 1]}\\
& $\theta_{S12}$ & 0 & 0  & & $M_{S12}$ & \\
& $\theta_{S13}$ & 0 & 0  & & $M_{S13}$ & \\
& $\vdots$ & $\vdots$ & $\vdots$  & & $\vdots$ & \\
& $\theta_{S21}$ & 0 & 0  & & $M_{S21}$ & \multirow{4}{*}[1.75em]{\rdelim\}{4}{*}[~Static case 2]}\\
& $\theta_{S22}$ & 0 & 0  & & $M_{S22}$ & \\
& $\theta_{S23}$ & 0 & 0  & & $M_{S23}$ & \\
& $\vdots$ & $\vdots$ & $\vdots$  & & $\vdots$ & \\
& \multicolumn{3}{c|}{$\cdots$} & & $\cdots$ & \hspace{0.9em}Static case 3 to $j$
\end{tabular}
\vspace{1em}
\caption{Training data structure before shuffling.}
\label{table.data}
\end{table}

After the $i=475$ dynamic cases and the $j=105$ static cases are collected, we assemble the $i+j=580$ cases together to construct the training dataset. A state vector containing four quantities - $\theta$, $\dot{\theta}$, $\ddot{\theta}$, and $M$ - is selected and joined together to form a four-column matrix representing every parameter combination. The $i+j=580$ cases are shuffled to test for the robustness and repeatability of the trained model (see Appendix). Table \ref{table.data} shows how the training dataset is assembled and structured before shuffling. The assembled data has 3 features - $\theta$, $\dot{\theta}$, and $\ddot{\theta}$ - as ``predictors'', and the aerodynamic moment, $M$, as the system ``response''. The final training dataset is a matrix of 17,029,725 rows and 4 columns. We want to emphasize again that only prescribed and static data (one-way coupling) are used as training data, no passive two-way coupling data are involved.

\subsection{Model training}{\label{sec.training}}

\begin{figure}
\centering
\includegraphics[width=.7\textwidth]{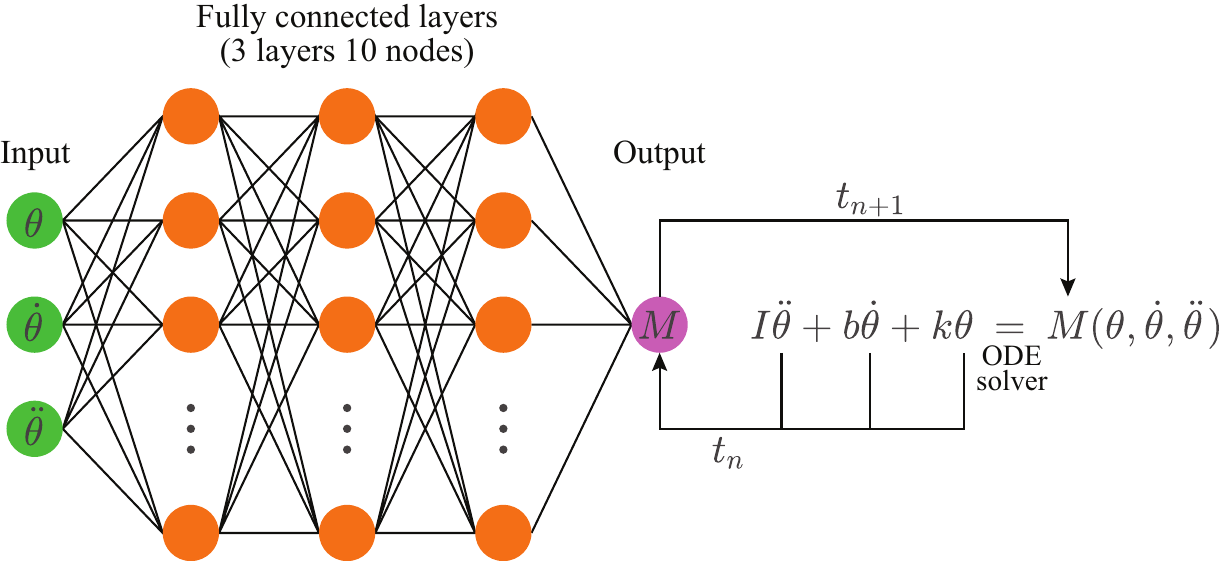}
\caption{A schematic showing the neural networks structure and how the trained model is integrated with the governing equation to simulate the instantaneous aeroelastic response.}
\label{fig.model}
\end{figure}

The training dataset is then fed into the MATLAB Regression Learner for training. In this study, we select feed-forward neural networks as our machine-learning regression model. We choose this model because it is known for being able to process data with high nonlinearity \citep{cybenko1989approximation}. The neural networks model has three fully connected layers with 10 nodes on each layer. Five-fold cross-validation is used to evaluate the predictive accuracy of the machine-learning model for one-way coupling cases. This validation method divides the dataset into five equal subsets (folds). The model is trained on four of these folds and tested on the remaining one. This process is repeated five times, with each fold being used as the test set once. The results from each iteration are averaged to give an overall performance estimate. This method reduces the bias and variance associated with random sampling, providing a more reliable measure of the model's ability to generalize to new data. The training time on an Intel\textregistered~Core\texttrademark~i7-14900HX CPU @ 2.2 GHz with 64 GB DDR5 memory without any GPU is about 5-6 hours. Fig. \ref{fig.model} shows a schematic of the neural networks structure used in the present study. We use three first-order features ($\theta$, $\dot{\theta}$, and $\ddot{\theta}$) to train the machine learning model for the aerodynamic moment $M=(\theta, \dot{\theta}, \ddot{\theta})$. In a separate test (not shown), we have included the second-order features (i.e. $\theta^2$, $\dot{\theta}^2$, $\ddot{\theta}^2$, $\theta\dot{\theta}$, $\theta\ddot{\theta}$ and $\dot{\theta}\ddot{\theta}$) and no significant improvements of the model performance were observed.

We want to emphasize that the goal of the present study is to propose and validate a data-driven approach for modeling nonlinear aeroelastic oscillations, rather than to design and optimize a specific machine-learning model. We chose the three-layered neural networks model because it is simple to implement yet complex enough to capture the essential nonlinearities in the training data.

\subsection{Aeroelastic response prediction and validation}{\label{sec.prediction}}

To make predictions for the aeroelastic response, we embed the trained aerodynamic moment model $M=(\theta, \dot{\theta}, \ddot{\theta})$ into the governing equation of the aeroelastic system (Eq. \ref{eqn.govern}). We use a customized Fourth Order Runge–Kutta (RK4) solver to solve the governing equation. The inertia, $I$, the structure damping, $b$, and the torsional stiffness, $k$, can be assigned to desired values to mimic physical experiments. For each combination of $I$, $b$, and $k$ (i.e. each simulation case), we solve Eq. \ref{eqn.govern} with a step size of 1/100 s and a duration of 120 seconds. At $t_0$, we assign initial values for $\theta_0$, $\dot{\theta}_0$ and $\ddot{\theta}_0$. Based on these initial values, we obtain the aerodynamic moment at the first time step, $M_1$, using the trained machine learning model. This $M_1$ is then fed into the RK4 solver to solve for $\theta_1$, $\dot{\theta}_1$ and $\ddot{\theta}_1$ at the next time step $t_1$. This loop continues until the time limit is reached (120 seconds for each case). To mimic our previous cyber-physical experiments \citep{zhu2020nonlinear}, the values of $\theta$, $\dot{\theta}$ and $\ddot{\theta}$ at the last time step of the current case are used as the initial condition for the next case. For each case, the kinematics of the system ($\theta$, $\dot{\theta}$ and $\ddot{\theta}$) as well as the aerodynamic moment ($M$) are collected as time series data for further analysis. This iteration loop is reflected on the right side of Fig. \ref{fig.model}.

In the present study, we fix the wing inertia, $I$, and the structural damping, $b$, and vary the torsional stiffness, $k$, resembling the bifurcation tests in physical experiments \citep{onoue2015large,zhu2020nonlinear,zhu2023flow}. Following previous experiments, we normalize the governing equation Eq. \ref{eqn.govern} using the fluid inertia to get the non-dimensional governing equation of the system:
\begin{equation}
    I^* \ddot{\theta}^* + b^* \dot{\theta}^* + k^* \theta^* = C_M,
\label{eqn.normlized_govern}
\end{equation}
where
\begin{equation}\label{eqn.parameters}
    \theta^* = \theta,~\dot{\theta}^*=\frac{\dot{\theta}c}{U_{\infty}},~\ddot{\theta}^* = \frac{\ddot{\theta}c^2}{U_{\infty}^2},~I^* = \frac{I}{0.5\rho c^4 s},~b^* = \frac{b}{0.5\rho U_{\infty} c^3 s},~k^* = \frac{k}{0.5\rho U_{\infty}^2 c^2 s},~C_M = \frac{M}{0.5\rho U_{\infty}^2 c^2 s}.
\end{equation}
Here $I^*$, $b^*$, $k^*$ and $C_M$ are non-dimensional inertia, structural damping, torsional stiffness and aerodynamic moment, respectively. In the ``virtual'' experiments, we use the inverse of the non-dimensional stiffness, or the Cauchy number, $Ca=1/k^*$, as the control parameter. The amplitude response, $A$, the frequency response $f_p$, and the moment response, $M$, are analyzed to quantify the dynamics of the virtual aeroelastic system. 

Results from these ``virtual'' experiments are compared against those obtained from physical two-way coupling aeroelastic experiments. These physical experiments are conducted using a force-feedback cyber-physical system (CPS, see the dash-line loop in Fig. \ref{fig.setup}a) \citep{zhu2020nonlinear,zhu2021nonlinear,zhu2023flow}. The CPS mimics the spring-mass-damper system using a high-bandwidth digital (``cyber'') control loop and at the same time measures the physical aerodynamic load using a force transducer. This type of system has been shown to be effective in studying a variety of fluid-structure interaction problems, and it is well known for its convenience in varying the structural parameters. The CPS allows us to control the mass of the wing and thus the mass ratio of the system (the ratio of the wing inertia to that of the surrounding fluid), and in this way the data acquired in a \emph{water} tunnel can be accurately applied to represent aeroelastic instabilities in \emph{air}. More details of the current CPS setup can be found in \citet{zhu2020nonlinear}.

\section{Results}{\label{sec.results}}
\subsection{Predictions of the system response - structural and hydrodynamic modes}

\citet{zhu2020nonlinear} discovered two distinct oscillation modes of aeroelastic pitching wings. For a high-inertia wing, the aeroelastic oscillations are dominated by the inertial force, corresponding to a ``structural mode'' in which high amplitude aeroelastic instability is characterized by a subcritical bifurcation and the oscillation frequency, $f_p^*=f_p c/U_\infty$, is locked into the structural frequency, $f_s^*=c\sqrt{k/I-(b/2I)^2}/(2\pi U_\infty)$. For a low-inertia wing, the aeroelastic oscillations are characterized by a supercritical bifurcation and the oscillation frequency is associated with vortex shedding at a constant Strouhal number, $St = f_p^* \approx 0.09$. They labeled this as a ``hydrodynamic mode''. To test if the proposed data-driven approach can accurately reproduce these two oscillation modes, we conduct ``virtual'' bifurcation tests and compare the results with those obtained from physical experiments \citep{zhu2020nonlinear,zhu2023flow}. We choose two wing inertia values, $I^*=10.6$ (structural mode) and $I^*=0.66$ (hydrodynamic mode), and fix the structural damping at a small value $b^*=0.13$. For each wing inertia, we start with a low Cauchy number, $Ca$ (i.e. high spring stiffness, $k^*$), and incrementally increase $Ca$ (decrease $k^*$) to test for the onset of large-amplitude limit-cycle oscillations (LCOs). We then decrease $Ca$ (increase $k^*$) to test for any hysteresis. At each $Ca$, time traces of the system dynamics ($\theta$, $\dot{\theta}$ and $\ddot{\theta}$) and the aerodynamic moment $M$ are generated using the data-driven approach. To mimic the physical experiments of \citet{zhu2020nonlinear}, the last value of the state vector for the current $Ca$ is used as the initial condition for the next $Ca$.

\begin{figure}[hbt!]
\centering
\includegraphics[width=1\textwidth]{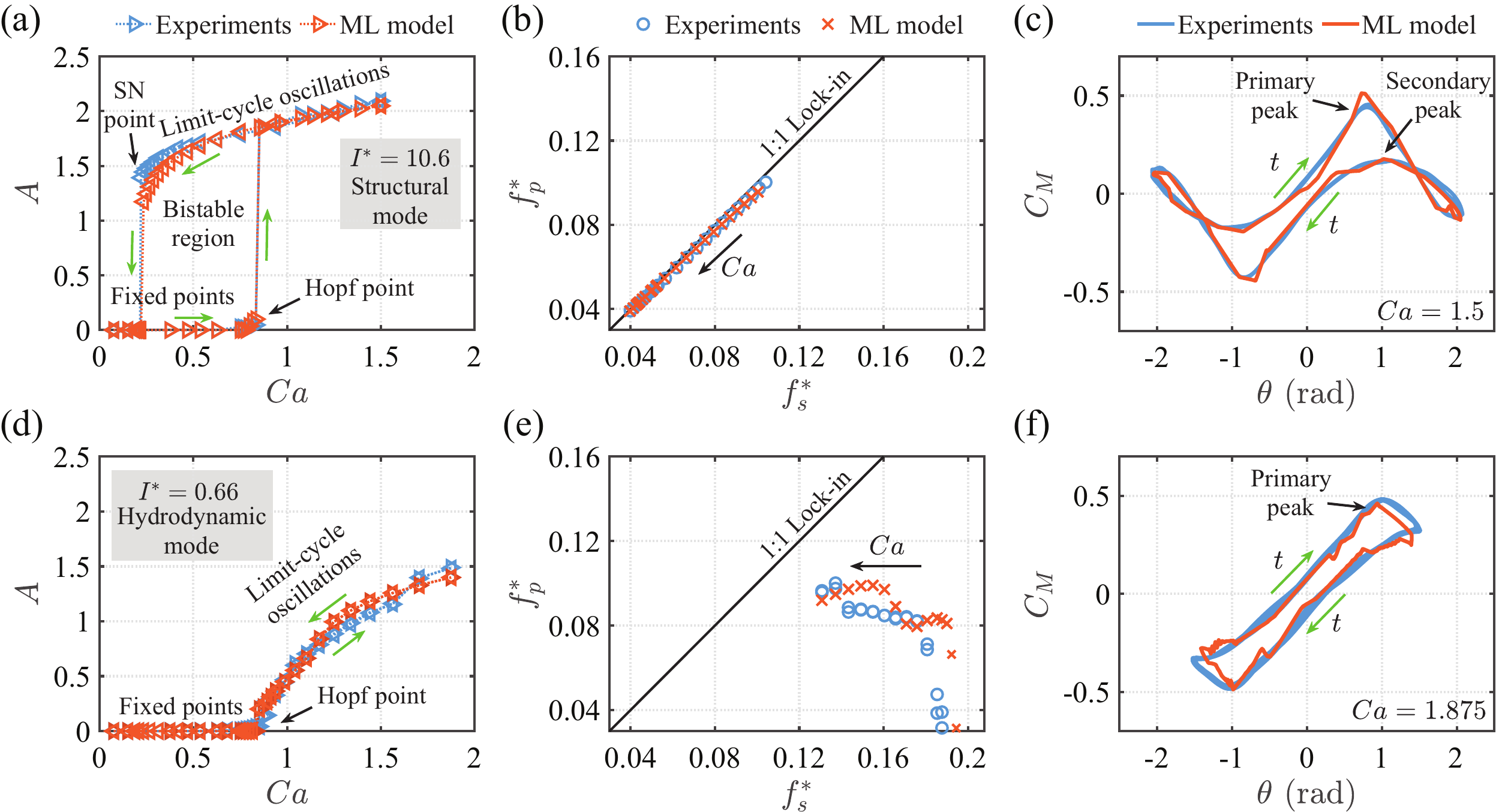}
\caption{Comparisons between results obtained from the proposed data-driven approach (orange markers) and physical experiments (blue markers). (a-c) A high-inertia wing ($I^*=10.6$) which corresponds to a \emph{structural} mode. (d-f) A low-inertia wing ($I^*=0.66$) which corresponds to a \emph{hydrodynamic} mode. (a,d) The amplitude response (bifurcation diagrams). (b,e) The frequency response. (c,f) The pitching angle-aerodynamic moment phase portrait.}
\label{fig.response}
\end{figure}

We plot the predicted amplitude response of the system, $A$ (i.e. the pitching amplitude, in radians), as a function of the control parameter, $Ca$, using orange markers in Fig. \ref{fig.response}(a) and (d). Results from the physical aeroelastic experiments are also plotted using blue markers for comparison. As shown in the figures, for both cases, the system first stays at a stable fixed point regime where no oscillations are observed due to the high spring stiffness. As we decrease the stiffness, static divergence and small-amplitude oscillations around the static divergence angle start to emerge. At a Hopf point, the system bifurcates to large-amplitude limit-cycle oscillations. For the high-inertia wing (Fig. \ref{fig.response}a), this bifurcation occurs abruptly and the system experiences a jump in amplitude, hallmarking a subcritical bifurcation \citep{strogatz1994nonlinear}. For the low-inertia wing (Fig. \ref{fig.response}b), however, the bifurcation is smooth and no amplitude jump is observed, indicating a supercritical bifurcation. For both cases, the amplitude of the oscillations increases as we keep increasing $Ca$ above the Hopf point. We then reverse the operation and decrease $Ca$. For the high-inertia case ($I^*=10.6$), the system stays at the large-amplitude LCO regime even when $Ca$ is decreased below the Hopf point. This hysteretic behavior creates a bistable region, which is another indicator of a subcritical bifurcation. Within the bistable region, the system has a stable fixed point solution, a stable LCO solution (i.e. bistability), and an unstable LCO solution (not observable in real experiments nor using the data-driven approach). The system finally falls back to the stable fixed point regime via a saddle-node (SN) point as $Ca$ is further decreased. In contrast, no hysteresis is observed for the low-inertia wing ($I^*=0.66$) and the system follows exactly the same route back to the stable fixed point regime - again, illustrating a hallmark of a supercritical bifurcation.

Comparing machine-learning results (orange) with experiments (blue), we see that the overall performance of the proposed data-driven approach is excellent. The machine-learning model accurately predicts the dynamical characteristics of the aeroelastic system for both the high-inertia structural mode and the low-inertia hydrodynamic mode. The two types of bifurcations (subcritical and supercritical) are faithfully reproduced with accurate prediction of the bifurcation points. The LCO amplitude of the machine-learning model also matches well with experiments, despite small discrepancies near the saddle-node point for the high-inertia case (Fig. \ref{fig.response}a). The averaged error of the predicted LCO amplitude is $2.48\%$ for the structural mode and $7.24\%$ or the hydrodynamic mode.

Fig. \ref{fig.response}(b) and (e) show the frequency response of the system, where the measured pitching frequency, $f_p^*$ is plotted against the natural (structural) frequency, $f_s^*$. For the high-inertia wing (Fig. \ref{fig.response}b), $f_p^*$ locks on to $f_s^*$, indicating a structural oscillation mode, where the oscillations are dominated by the inertia force \citep{zhu2020nonlinear}. The pitching frequency predicted by the data-driven approach matches very well with experiments. The slight deviation between the frequency response and the 1:1 lock-in line, which is presumably caused by the added-mass effect \citep{zhu2020nonlinear}, is also accurately reflected in the machine-learning model. Since the full aerodynamic moment ($M$) was used as the training data, this includes the added-mass contributions, thus this effect is incorporated into the neural networks predictions. For the low-inertia wing (Fig. \ref{fig.response}e), as we increase $Ca$, the measured pitching frequency $f_p$ remains relatively constant and stays below the 1:1 lock-in value. This shows that the oscillations are dictated by a fluid time scale and thus the fluid force, hallmarking a hydrodynamic mode \citep{zhu2020nonlinear}. The machine-learning model faithfully captures the frequency response of this mode, despite a higher discrepancy as compared to the prediction of the structural mode. The averaged error of the predicted pitching frequency is $0.41\%$ for the structural mode and $16.18\%$ or the hydrodynamic mode. The structural mode has a much more accurate predicted frequency because in the high-inertia case, the frequency of the motion is driven by the competing inertial force and spring restoring force, while the unsteady aerodynamic forcing plays a secondary role (see Fig. 4b of \citet{zhu2020nonlinear}). Therefore, the oscillation frequency is mostly determined by the structural forces (which also explains the lock-in phenomenon). In the low-inertia case, the oscillations are driven by the competing aerodynamic force and the spring restoring force, making the frequency more sensitive to the accuracy of the predicted aerodynamic forcing term.

Finally, the moment responses of the system (i.e. the moment-angle phase portrait) at the maximum Cauchy number ($Ca=1.5$ for $I^*=10.6$ and $Ca=1.875$ for $I^*=0.66$) are shown in Fig. \ref{fig.response}(c) and (f). We see that the moment response predicted by the data-driven method agrees well with the experimental results. \citet{zhu2020nonlinear} discussed the interpretation of these phase maps and their relation to the vortex dynamics in detail. For the high-inertia wing (Fig. \ref{fig.response}c), the primary peak in the moment induced by the formation and shedding of a strong leading-edge vortex (LEV) during the upstroke, as well as the secondary peak induced by a secondary LEV during the pitch reversal, are both accurately captured by the data-driven method. For the low-inertia wing (Fig. \ref{fig.response}f), the wing only sheds one LEV during the upstroke. Therefore, only one peak is observed in the phase portrait. Fig. \ref{fig.response}(c) and (f) show that the pitching aerodynamic moment is predicted accurately in a two-way coupling response despite that the machine-learning model does not include training cases for arbitrary prescribed motions (i.e., non-sinusoidal) or aeroelastic motions. The accurate prediction of the aerodynamic moment enables good agreement between the data-driven and the experimental results in the amplitude response (Fig. \ref{fig.response}a,d) and the frequency response (Fig. \ref{fig.response}b,e). It is also worth mentioning that our machine-learning model demonstrates that by using the time series data of only one parameter - the pitching angle (as the pitching rate and accelerations are time derivatives of the pitching angle) - we are able to make fairly accurate predictions of the highly nonlinear and complex unsteady aerodynamic loads for elastically mounted wings undergoing (near) sinusoidal pitching motion in vortex-dominated flows.

\subsection{Predictions of the unsteady transition dynamics}

\begin{figure}
\centering
\includegraphics[width=0.8\textwidth]{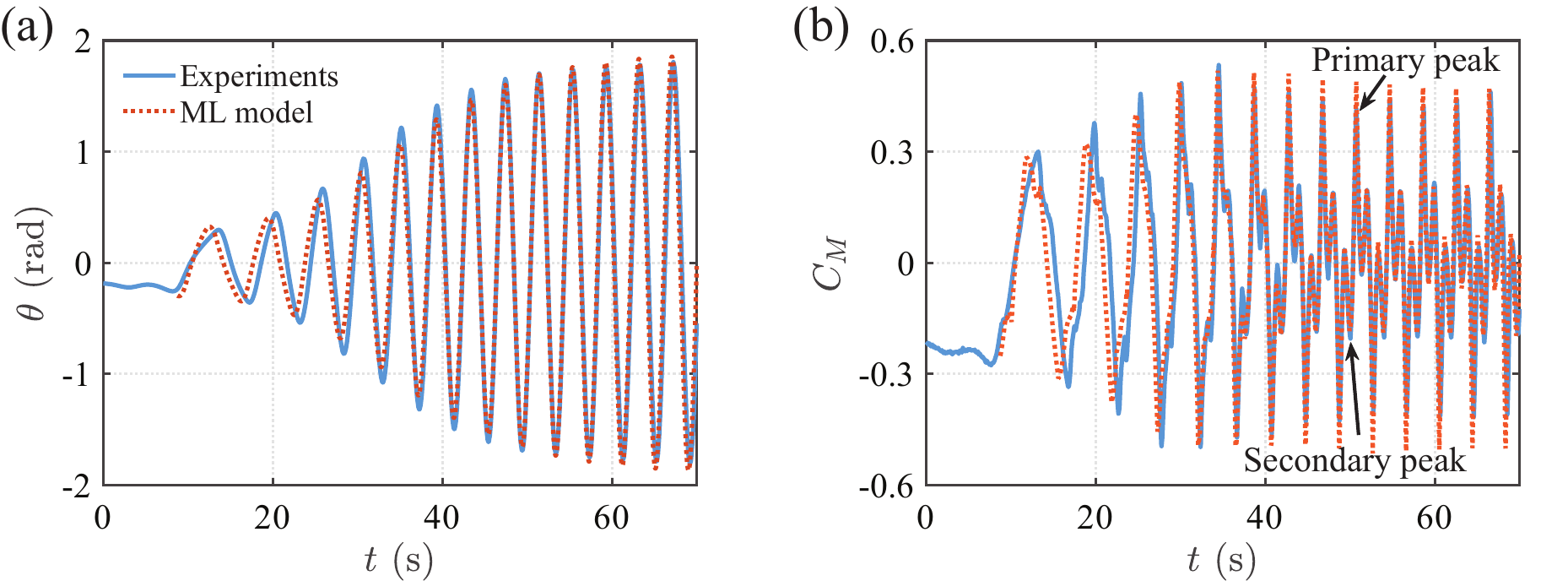}
\caption{(a) Pitching amplitude transition to large-amplitude limit-cycle oscillations right after the Hopf point for the case $I^*=10.6$, $b^*=0.13$, $Ca=0.85$. (b) The corresponding evolution of the aerodynamic moment.}
\label{fig.transition}
\end{figure}

Another important feature of the aeroelastic system dynamics is the transition of the system from a stable fixed point to large-amplitude limit-cycle oscillations, which occurs as the Cauchy number first rises above its critical value at the bifurcation point (Fig. \ref{fig.response}a,d). In particular, it is of great practical importance to characterize the transition after the subcritical Hopf point (Fig. \ref{fig.response}a), because the subcritical bifurcation is abrupt and often considered to be dangerous in engineering applications \citep{strogatz1994nonlinear}. Right after the bifurcation point, the system experiences a rapid amplitude growth before saturating into large-amplitude limit-cycle oscillations. 

To test if the proposed data-driven method can accurately capture this transition, in Fig. \ref{fig.transition}, we compare the machine-learning-predicted pitching amplitude and aerodynamic moment as a function of time, with those obtained from experiments. The transition occurs at $I^*=10.6,~b^*=0.13,~Ca=0.85$. For better comparisons, the offset of the predicted time trace is adjusted to align with the experimental results. We see that both the amplitude and the aerodynamic moment evolution predicted by the data-driven approach match well with experiments. This agreement is excellent considering that we have only used prescribed \emph{periodic} oscillations with \emph{constant} amplitudes and static measurements at \emph{fixed} angles of attack for training the machine learning model; no transition data are included in the training dataset (Table \ref{table.data}). \citet{zhu2020nonlinear} reported the emergence of a secondary LEV during the transition to large-amplitude LCOs, which corresponds to an emerging secondary peak in the aerodynamic moment measurement. This process is also faithfully captured by the data-driven method, as depicted in Fig. \ref{fig.transition}(b). We can see that the amplitude of the secondary peak grows and saturates during the transition. Fig. \ref{fig.transition} shows that the machine learning model is able to make good predictions for unsteady dynamics even though it is trained only using steady-state, harmonic data. We believe this success is due to the fact that the time scale associated with the state transition is slow as compared to the time scale of the limit-cycle oscillations. This result is consistent with our previous ``ring-down'' experiments \citep{zhu2021nonlinear} where we have shown that cycle-averaged values can be used to quantify the slow decay of oscillations due to fluid damping if the transition time scale is slow.

\subsection{Performance of the machine-learning model over a range of parameters}

\begin{figure}
\centering
\includegraphics[width=1\textwidth]{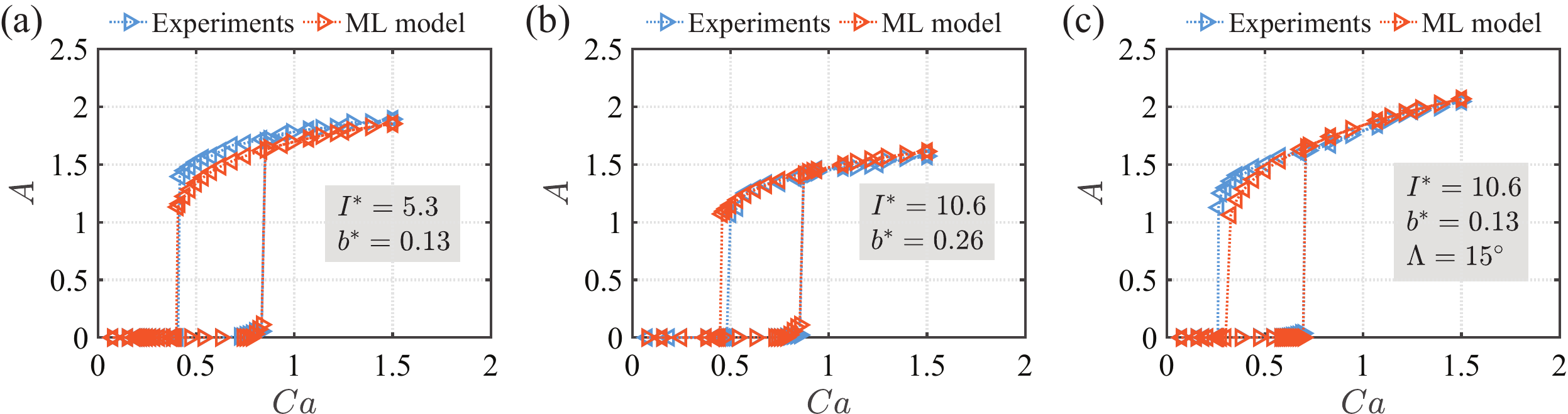}
\caption{Bifurcation diagram comparisons between experiments and the data-driven approach for additional cases. (a) $I^*=5.3$ and $b^*=0.13$ (half the inertia of Fig. \ref{fig.response}a). (b) $I^*=10.6$ and $b^*=0.26$ (double the damping of Fig. \ref{fig.response}a). (c) $I^*=10.6$ and $b^*=0.13$ with a leading-edge sweep angle of $\Lambda=15^\circ$ (requires training a new neural networks model with data taken at $\Lambda=15^\circ$).}
\label{fig.additional}
\end{figure}

To demonstrate the broad applicability of the proposed data-driven approach, we test the machine-learning model on three additional cases, and compare the results with physical experiments in Fig. \ref{fig.additional}. In Fig. \ref{fig.additional}(a), we reduce the wing inertia of case Fig. \ref{fig.response}(a) to half, which results in $I^*=5.3$ and $b^*=0.13$. We see that at this wing inertia, the system still undergoes a subcritical Hopf bifurcation to large-amplitude LCOs, despite a smaller bistable region. The machine-learning model precisely predicts both the Hopf point and the saddle-node point, although it slightly underpredicts the LCO amplitude. In Fig. \ref{fig.additional}(b), we maintain the same wing inertia as in Fig. \ref{fig.response}(a), $I^*=10.6$, and double the structural damping to $b^*=0.26$. The LCO amplitude becomes lower at this damping value, but the system still features a subcritical bifurcation and a structural oscillation mode. For this case, the LCO amplitude is precisely predicted by the machine-learning model, but the predicted saddle-node point has a slight discrepancy as compared to experiments. Note that in order to run the cases of Fig. \ref{fig.additional}(a) and (b), we only need to change the structural parameters in the governing equation (see step 2 in Fig.\ref{fig.setup}b and \S \ref{sec.prediction}). The trained machine-learning model stays the same and no additional data or training is required.

The machine-learning model does have limitations. We note that this machine-learning model does not reproduce the experimental results for extremely small values of wing inertia. For example, we have tested the model with $I^*=0.33$ (results not shown), and large-amplitude LCOs are not observed for the $Ca$ range tested except at the maximum value, $Ca=1.875$. Instead, the model predicts the system to have small-amplitude oscillations ($A<0.3$). The machine-learning model does not allow us to further investigate the cause of this discrepancy because it is non-interpretable. However, we hypothesize that as the wing inertia becomes too small, the unsteady fluid forcing term becomes dominant, and the timing and the dynamics of the aeroelastic system become very sensitive to the exact phase and amplitude of the fluid forcing. Adding additional higher-order predictors to the model does not fix the issue. Although our current neural networks model is good enough to capture the complex dynamics of both the structural mode and the hydrodynamic mode (Fig. \ref{fig.response}), more sophisticated machine-learning models may be helpful for predicting the system dynamics of extremely low-inertia wings.

In Fig. \ref{fig.additional}(c), we extend the data-driven approach to \emph{swept} wings. A detailed description of the swept wing geometry and the characteristics of the associated nonlinear instabilities is reported by \citet{zhu2023flow}. Here we take the swept wing with a sweep angle $\Lambda=15^\circ$ as an example. The wing inertia and damping are maintained at $I^*=10.6$ and $b^*=0.13$. Because the wing geometry has changed, all the training data has to be retaken and the machine-learning model has to be retrained following the procedures discussed in \S \ref{sec.data_prepare} and \S \ref{sec.training}. The Hopf point as well as the LCO amplitude after the Hopf point are precisely predicted. The machine-learning model slightly underpredicts the LCO amplitude in the bistable region, and the saddle-node point is predicted to be at a slightly higher $Ca$ as compared to experiments. Overall, the machine-learning model is able to predict the system dynamics well for the swept wing, demonstrating the wide applicability of the proposed approach.

\subsection{Potential scalability of the machine-learning model}

\begin{figure}
\centering
\includegraphics[width=0.8\textwidth]{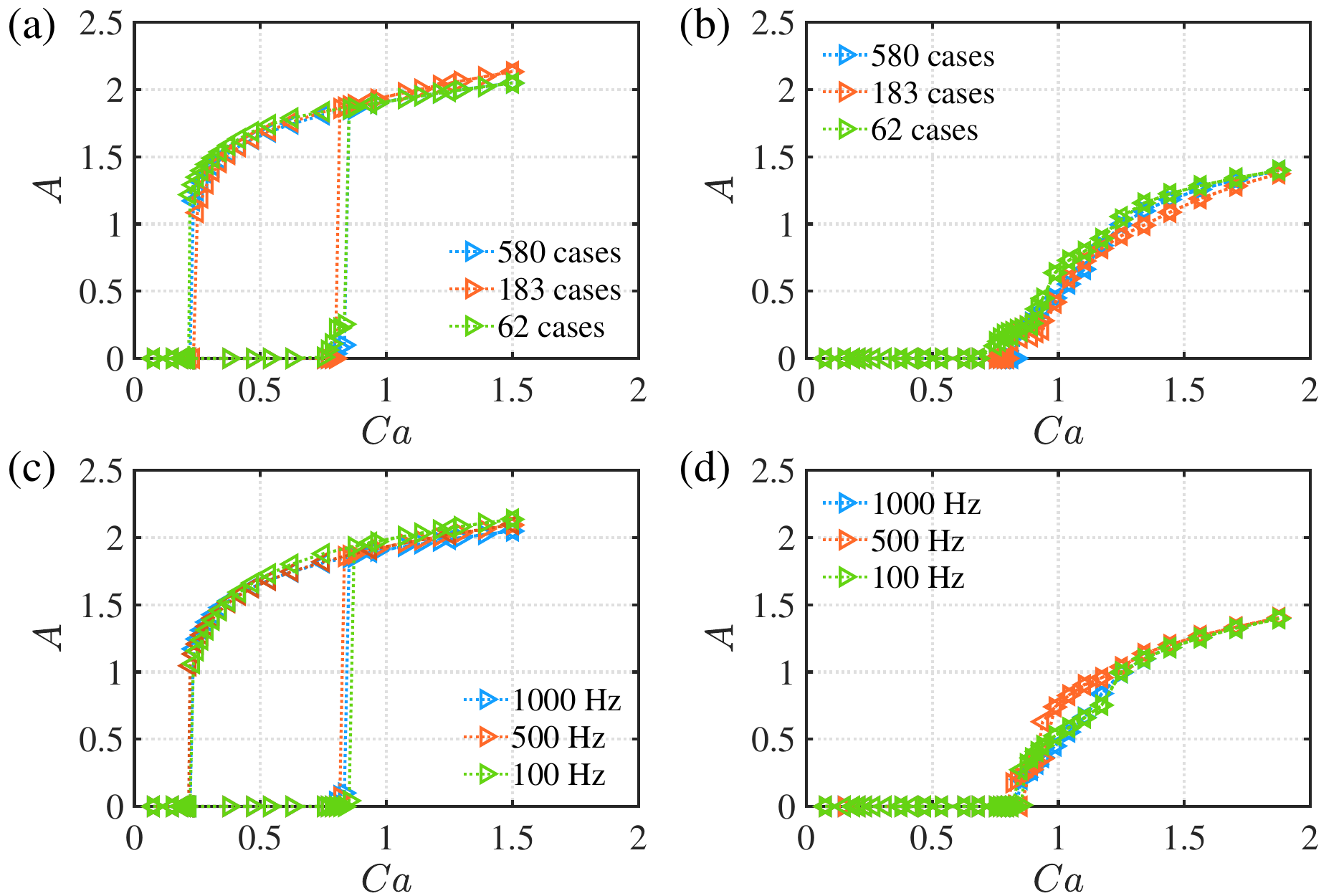}
\caption{(a,b) Results from the proposed data-driven approach using 580, 183, and 62 training cases. (c,d) Results from the proposed data-driven approach using sampling rates of 1000 Hz, 500 Hz, and 100 Hz for the training data. (a,c) Inertia-dominated case ($I^*=10.6$), corresponding to a \emph{structural} mode. (b,d) Fluid-dominated case ($I^*=0.66$), corresponding to a \emph{hydrodynamic} mode.}
\label{fig.scalability}
\end{figure}

In this study, we developed a data-driven approach for modeling a 1-DOF aeroelastic system. To scale this model up for more realistic aeroelastic systems with a higher number of DOFs, it is important to examine the effect of training dataset size. In previous sections, we used 580 training cases with a sampling rate of 1,000 Hz per case to construct the $17,029,725\times4$ training dataset (see \S \ref{sec.method}). This large dataset may present challenges in scaling the model for systems with more DOFs, particularly when using experimental or high-fidelity CFD data. Therefore, we reconstruct the training dataset with fewer cases and lower sampling rates, and retrain the machine learning model.

Fig. \ref{fig.scalability} shows the results of the retrained models. In Fig. \ref{fig.scalability}(a,b), we test the effect of the number of training cases on (a) the structural mode and (b) the hydrodynamic mode. The dataset with 580 cases corresponds to the dataset used in previous sections, consisting of $i=475$ dynamic cases and $j=105$ static cases ($17,029,725\times4$ total data size). We reduce the number of cases by skipping every other dynamic and static case, resulting in $i=130$ dynamic cases and $j=53$ static cases, for a total of $i+j=183$ cases. Note that for the dynamic cases, we skip every other case in both the frequency $f-$dimension and the amplitude $A-$dimension, see Fig. \ref{fig.data}(a). The size of this dataset is reduced to $5,229,490\times4$, around 31\% of the original data size. To further reduce the case number, we skip every other case again, obtaining a new dataset with only $i+j=62$ cases, where $i=35$ and $j=27$. The size of this dataset is $1,769,795\times4$, only 10\% of the original data size. For all the three cases ($i+j=580$, $i+j=183$, and $i+j=62$), the sampling rate was kept at 1000 Hz. As shown in Fig. \ref{fig.scalability}(a,b), although the training dataset is reduced to only 62 cases, our approach is still capable of capturing both subcritical and supercritical bifurcations in the structural and hydrodynamic modes. The predicted oscillation amplitudes for the 580-case model, the 183-case model, and the 62-case model are very close, with only minor deviations.

Another way to reduce the training data size is by lowering the sampling rate (i.e., downsampling the training data). This effect is shown in Fig. \ref{fig.scalability}(c, d). For a case number of 580, three sampling rates are considered: 1000 Hz, which corresponds to the rate used in previous sections, and the reduced rates of 500 Hz and 100 Hz. The dataset with a 500 Hz sampling rate has a total size of $8,514,863\times4$, 50\% of the original data size, while the dataset with a 100 Hz sampling rate has a total size of $1,702,973\times4$, only 10\% of the original data size. Similar to the reduced case number models, the models with reduced sampling rates are still able to capture the essential bifurcation features, with only small deviations from the original 1000 Hz model. Overall, the predictions remain satisfactory.

Further reducing the case number or sampling frequency leads to numerical instabilities in the aeroelastic model (i.e. the amplitude grows exponentially to extremely high magnitude). Although we do not have an exact explanation for this numerical instability, we note that it only occurs in the two-way coupling aeroelastic simulation, whereas the one-way coupling aerodynamic moment still shows satisfactory cross-validation results in the machine learning training. This indicates that the numerical instability is associated with additional down-sampling, which skips frequencies close to those of the physical aeroelastic phenomena or uses sampling rates that fail to capture the relevant frequencies (leading to aliasing or chopped oscillations). For example, for the fluid-dominated case ($I^*=0.66$) in Fig. \ref{fig.scalability}(b,d), the LCO frequency is around $f^*=0.1$ (see Fig. \ref{fig.response}e). Reducing the case number from 580 to 62 leaves the pitching frequencies of the training data at $f^*=0.03, 0.05, 0.07, 0.09$ and $0.11$, which includes two frequencies ($f^*=0.09$ and $0.11$) close to $f^*=0.1$. However, further reducing the number of cases results in no training data near the physical aeroelastic oscillation frequency, leading to potential numerical instabilities. There may exist an optimal combination of case number and sampling rate that minimizes the total size of the training dataset while maintaining a high level of prediction accuracy. This optimal combination should depend on the retained frequencies of the sinusoidal training dataset and the sampling rate, since one needs an appropriate sampling rate to accurately resolve a given frequency. However, identifying that optimum is beyond the scope of the present study. Nevertheless, the models with reduced case numbers and reduced sampling frequencies demonstrate that our approach has good scalability potential, suggesting it could be applied to systems with higher DOFs.

\section{Discussions and Conclusion}{\label{sec.conclusion}}

In this study, we proposed and validated a data-driven approach for modeling large-amplitude flow-induced oscillations of elastically mounted pitching wings with full flow nonlinearity and at a low computational cost. To achieve this goal, we trained a neural networks regression model to predict the nonlinear aerodynamic moment experienced by the wing based on the wing kinematics, namely the pitching angle, rate, and acceleration. The training dataset was generated purely by one-way coupling prescribed motions and static measurements without any prior knowledge of the two-way coupling passive aeroelastic dynamics. The trained model was then combined with the aeroelastic governing equation and fed into an ordinary differential equation solver to simulate the passive aeroelastic system. ``Virtual'' experiments of high- and low-inertia wings were performed using the proposed method and the results were compared with physical experiments. The amplitude, frequency, and aerodynamic moment responses of the system were found to be well predicted by the proposed method, with the structural and hydrodynamic oscillation modes faithfully recreated. The unsteady growth and saturation of the pitching amplitude and the aerodynamic moment right after the bifurcation point were also successfully captured. Validations of the proposed approach on additional test cases demonstrated its high versatility. Due to the data-driven nature of this approach, the specific geometry of the elastically supported object does not come into play in the model, making this approach broadly applicable and easily adaptable to other aeroelastic systems.

The proposed data-driven approach successfully bridges one-way and two-way coupling fluid-structure interaction problems, opening up the possibility of using lower-cost, more accessible one-way coupling experiments/simulations to study higher-cost, less accessible two-way coupling FSI problems. More importantly and perhaps more excitingly, following the proposed approach, the excessive existing datasets of unsteady airfoils/wings with prescribed kinematics in the literature can potentially be consolidated and used to train a more comprehensive data-driven model for studying more complex and realistic aeroelastic problems. The proposed approach also echoes the concept of ``digital twins'' \citep{jones2020characterising}, in a way that a ``digital'' aeroelastic system is created and trained using data obtained from its physical counterpart. And this ``digital twin'' can in turn be used to simulate and guide the study of the physical aeroelastic system.


We want to emphasize again that the goal of the present study is to propose and validate a data-driven approach for modeling nonlinear aeroelastic systems, rather than to design and optimize a specific machine-learning model. We chose the three-layered neural networks model because it is simple to implement yet complex enough to capture the essential nonlinear dynamics of the unsteady aerodynamic moment. In fact, it is quite amazing that the highly nonlinear aerodynamic moment of pitching wings in massively-separated, vortex-dominated flows can be accurately predicted by the three very basic wing kinematics: pitching angle, rate, and acceleration, without additional information about surface pressure or velocity field. Unfortunately, because the neural networks model is non-interpretable, it is difficult to infer any useful information from the model. Employing more interpretable machine-learning models such as SINDy \citep{brunton2016discovering} might be able to provide more physical insights, but this lies out of the scope of the present study. 

At the current stage, the proposed data-driven approach is not able to predict flow-induced oscillations of extremely low-inertia wings, although the high-inertia structural mode and the low-inertia hydrodynamic mode can both be accurately predicted. It is hypothesized that for extremely low-inertia wings, the system becomes too sensitive to the exact phase and amplitude of the fluid forcing. In addition, the neural networks model used in the proposed approach is a regression model, which means it is good at interpolating data. How good it is in extrapolating data (i.e. predicting aeroelastic dynamics beyond the amplitude and frequency range of the training data) remains a question for future studies. Moreover, the current work was only performed at a single Reynolds number, and for a particular wing geometry. Although we have previously shown that these nonlinear aeroelastic instabilities are generally insensitive to Reynolds number \citep{zhu2020nonlinear}, these effects are nevertheless important, and one would expect that the predictions of the fluid moment, and the details of the bifurcation points and oscillation amplitudes will change with Reynolds number, and thus will likely require additional input data and model training. Lastly, the present study only considered an idealized 1-DOF aeroelastic system. Real aeroelastic systems often have flexible wings that feature a much larger number of DOFs. Although we have demonstrated good scalability potential of the model, it is important to test the proposed model in more realistic systems in the future.

\section*{CRediT authorship contribution statement}
\textbf{Yuanhang Zhu:} Conceptualization, Formal Analysis, Investigation, Methodology, Software, Validation, Visualization, Writing – original draft, Writing – review \& editing. \textbf{Kenneth Breuer:} Conceptualization, Formal Analysis, Investigation, Methodology, Writing – review \& editing.

\section*{Declaration of competing interest}
The authors declare that they have no known competing financial interests or personal relationships that could have appeared to influence the work reported in this paper.

\section*{Data availability}
Data will be made available on request.

\section*{Acknowledgments}
This work was funded by the Air Force Office of Scientific Research, Grant FA9550-21-1-0462, managed by Dr. Gregg Abate.

\appendix
\section{Robustness test of the model}{\label{sec.repeatability}}

\begin{figure}[ht!]
\centering
\includegraphics[width=1\textwidth]{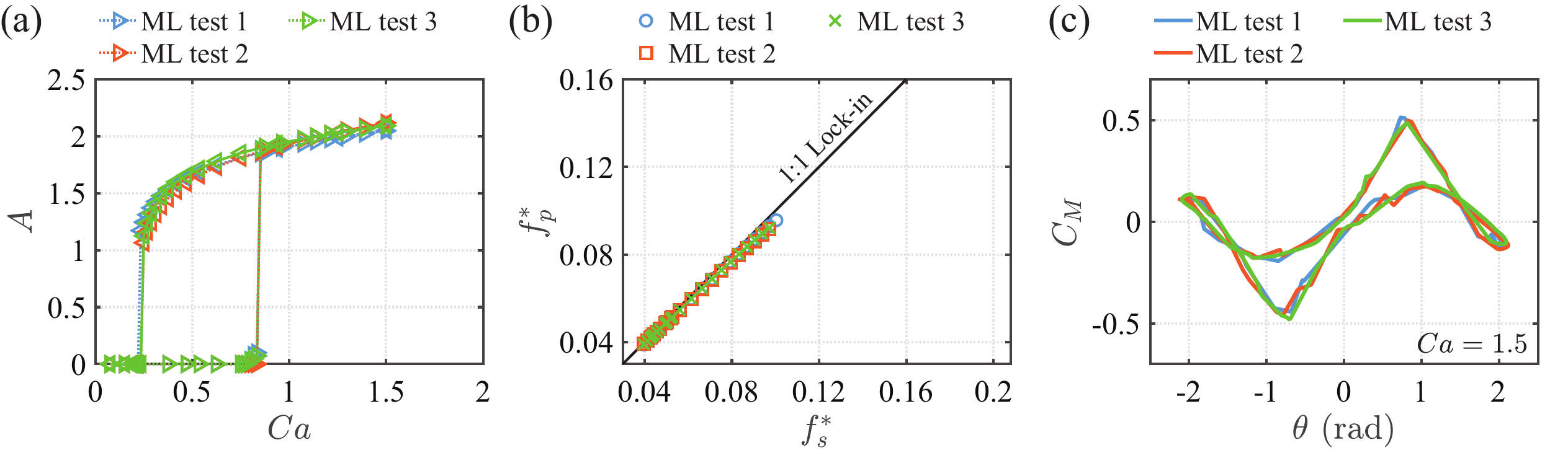}
\caption{Robustness test of the model using the high-inertia case, $I^*=10.6$ and $b^*=0.13$. The models are trained using the same neural networks structure but the training data is shuffled differently.}
\label{fig.repeat}
\end{figure}

To test the robustness and repeatability of the machine learning model, we repeat the high-inertia case of Fig. \ref{fig.response}(a-c). We reshuffle the 580 dynamic and static cases of the training data (see \S \ref{sec.data_prepare}) and use the reordered data to train the neural networks model. Note that shuffling changes the order of the training cases while keeping the data for each case in sequence. We shuffle the cases so that the MATLAB Regression Learner treats each shuffled dataset as a new dataset. After retraining the model, we repeat the bifurcation tests, and the results are shown in Fig. \ref{fig.repeat}. The reshuffling and the retraining are repeated twice, corresponding to machine-learning tests 2 and 3 in the figure. The results of Fig. \ref{fig.response}(a-c) are replotted as ML test 1. We see that the amplitude, frequency, and aerodynamic moment response of the three machine-learning tests are highly repeatable despite some small variations, demonstrating the high robustness and repeatability of the model.

\section{Effect of noise on the model}{\label{sec.noise}}

\begin{figure}[ht!]
\centering
\includegraphics[width=0.8\textwidth]{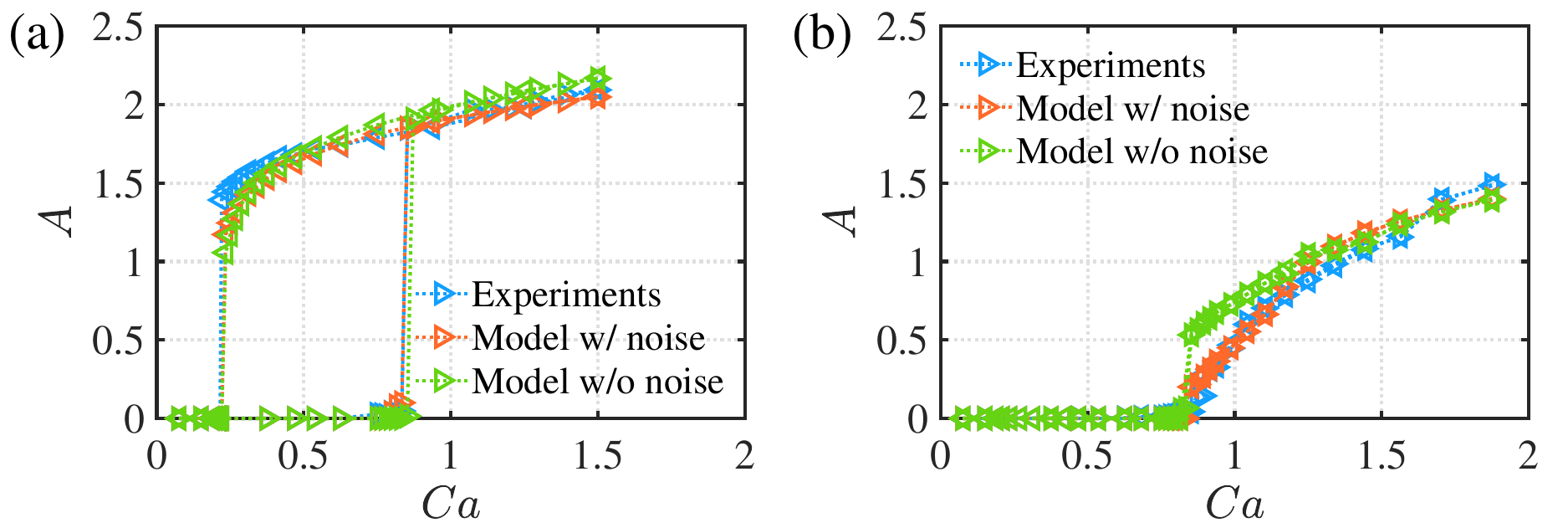}
\caption{Effect of noise filtering on the model performance. (a) Structural mode ($I^*=10.6$), and (b) Hydrodynamic mode ($I^*=0.66$).}
\label{fig.noise}
\end{figure}

As mentioned in \S \ref{sec.data_prepare}, we used raw aerodynamic moment data without noise filtering to construct the training dataset. It is both relevant and important to study the effect of noise on the model’s performance to assess whether a machine-learning model trained on CFD data would miss any important physics of the experiment. To filter the raw data for the dynamic cases, we apply a zero-phase, sixth-order low-pass Butterworth filter with a cutoff frequency set at 15 times the pitching frequency. For the static cases, the mean value of the 20-second raw data is calculated and used to replace the raw static aerodynamic moments. The machine learning model is then retrained using the filtered data and compared to the previous model, which was trained using the raw data. The comparison is shown in Fig. \ref{fig.noise}. We can see that the model with noise makes predictions closer to the experimental data, especially for the supercritical bifurcations of the hydrodynamic mode. This is not surprising, as the inherent noise in the aerodynamic moment is part of the forcing term in the aeroelastic system (Eq. \ref{eqn.govern}) and plays an important role in shaping the overall aeroelastic response. The noise filtering of the aerodynamic moment seems to have a more significant effect on the aeroelastic prediction of the hydrodynamic mode ($I^*=0.66$, Fig. \ref{fig.noise}b) than the structural mode ($I^*=10.6$, Fig. \ref{fig.noise}a), presumably because the hydrodynamic mode is primarily driven by the fluid force (i.e. the aerodynamic moment) while the structural mode is primarily driven by the inertia \citep{zhu2020nonlinear}.

Despite higher discrepancies in the aeroelastic predictions with noise filtering, the overall performance of the proposed approach is still acceptable. Inevitably, aeroelastic systems simulated using high-fidelity CFD exhibit slightly different behaviors from experimental data \citep{fagley2024leading}. Our results show that the proposed approach could be applied using low-noise, high-fidelity CFD data as training datasets. Furthermore, one could integrate experimental data and high-fidelity CFD data to predict the aeroelastic behavior, potentially allowing for a larger dataset.

\bibliographystyle{elsarticle-harv}
\bibliography{ref}

\end{document}